\RequirePackage{lineno}
\documentclass[twocolumn,showpacs,aps,prd,superscriptaddress,floatfix]{revtex4-1}
\usepackage{graphicx}  
\usepackage{dcolumn}   
\usepackage{bm}        
\usepackage{amssymb}   
\usepackage{amsfonts}  
\usepackage{hyperref}
\usepackage{amsmath}
\usepackage{mathtools}
\usepackage{xspace}
\usepackage{color}
\usepackage{mathrsfs}

\begin{document}
\title{\bf  Search for a \boldmath{$CP$}-odd light Higgs boson in \boldmath{$J/\psi \to \gamma A^0$ }}

\author{
		\begin{small}
		\begin{center}
M.~Ablikim$^{1}$, M.~N.~Achasov$^{10,b}$, P.~Adlarson$^{68}$, S. ~Ahmed$^{14}$, M.~Albrecht$^{4}$, R.~Aliberti$^{28}$, A.~Amoroso$^{67A,67C}$, M.~R.~An$^{32}$, Q.~An$^{64,50}$, X.~H.~Bai$^{58}$
% ...
\\
		\vspace{0.2cm}
		(BESIII Collaboration)\\
		\vspace{0.2cm} {\it
$^{1}$ Institute of High Energy Physics, Beijing 100049, People's Republic of China\\
% ...
$^{4}$ Bochum Ruhr-University, D-44780 Bochum, Germany\\
% ...
$^{10}$ G.I. Budker Institute of Nuclear Physics SB RAS (BINP), Novosibirsk 630090, Russia\\
% ...
$^{14}$ Helmholtz Institute Mainz, Staudinger Weg 18, D-55099 Mainz, Germany\\
% ...
$^{28}$ Johannes Gutenberg University of Mainz, Johann-Joachim-Becher-Weg 45, D-55099 Mainz, Germany\\
% ...
$^{32}$ Liaoning Normal University, Dalian 116029, People's Republic of China\\
% ...
$^{50}$ State Key Laboratory of Particle Detection and Electronics, Beijing 100049, Hefei 230026, People's Republic of China\\
% ...
$^{58}$ University of Jinan, Jinan 250022, People's Republic of China\\
% ...
$^{64}$ University of Science and Technology of China, Hefei 230026, People's Republic of China\\
% ...
$^{67}$ University of Turin and INFN, (A)University of Turin, I-10125, Turin, Italy; (B)University of Eastern Piedmont, I-15121, Alessandria, Italy; (C)INFN, I-10125, Turin, Italy\\
$^{68}$ Uppsala University, Box 516, SE-75120 Uppsala, Sweden\\
% ...
$^{b}$ Also at the Novosibirsk State University, Novosibirsk, 630090, Russia\\
% ...
}\end{center}
		\vspace{0.4cm}
		\end{small}
}

\begin{abstract} Using $J/\psi$ radiative decays from 9.0 billion
$J/\psi$ events collected by the BESIII detector, we search for
di-muon decays of a $CP$-odd light Higgs boson ($A^0$), predicted by
many new physics models beyond the Standard Model, including the
Next-to-Minimal Supersymmetric Standard Model. No evidence for the $CP$-odd
light Higgs production is found, and we set $90\%$ confidence level upper
limits on the product branching fraction $\mathcal{B}(J/\psi \to \gamma A^0)\times
\mathcal{B}(A^0 \to \mu^+\mu^-)$ in the range of $(1.2-778.0)\times
10^{-9}$ for $0.212 \le m_{A^0} \le 3.0$ GeV/$c^2$. The new
measurement is a 6-7 times improvement over our previous measurement, 
and is also slightly better than the BaBar measurement in the low-mass
region for $\tan \beta =1$.

\end{abstract}

\maketitle 

The origin of mass is one of the most important questions in physics.
The masses of the fundamental particles are generated through
spontaneous breaking of electroweak symmetry by the Higgs
mechanism~\cite{higgs}. The Higgs mechanism implies the existence of
at least one new scalar particle, the Higgs boson, which was the last
missing Standard Model (SM) particle. It was discovered by the Large
Hadron Collider experiments at CERN~\cite{Aad} in July 2012 and has a
profound effect on our fundamental understanding of matter.

Many models beyond the SM, such as the Next-to-Minimal Supersymmetric
Standard Model (NMSSM)~\cite{Hiller, Dermisek, Steggermann}, extend
the Higgs sector to include additional Higgs fields. The NMSSM adds an
additional singlet chiral superfield to the Minimal Supersymmetric
Standard Model (MSSM)~\cite{MSSM} to alleviate the so-called
\rm{\lq\lq little hierarchy problem\rq\rq}~\cite{Delgado}. It contains
three $CP$-even, two $CP$-odd, and two charged Higgs
bosons~\cite{Hiller, Dermisek}. The mass of the lightest Higgs boson,
$A^0$, may be smaller than twice the mass of the charmed quark,
thus making it accessible via $J/\psi \to \gamma A^0$~\cite{Wilczek}.

The branching fraction of $V \to \gamma A^0$ ($V=\Upsilon, J/\psi$) is
expressed as~\cite{Wilczek, Mangano, Nason} \begin{equation}
\frac{\mathcal{B}(V\to \gamma A^0)}{\mathcal{B}(V\to
l^+l^-)}=\frac{G_Fm_q^2g_q^2C_{\rm QCD}}{\sqrt{2}\pi
\alpha}\left(1-\frac{m_{A^0}^2}{m_V^2}\right), \label{yukawa}
\end{equation} \noindent where $\alpha$ is the fine structure
constant, $G_F$ is the Fermi coupling constant, $l \equiv e$ or $\mu$, $m_q$ is the quark mass, $C_{\rm
QCD}$ includes the leptonic width of $\mathcal{B}(V \to
l^+l^-)$~\cite{Barbieri,Beneke} as well as $m_{A^0}$ dependent QCD and
relativistic corrections to $\mathcal{B}(V \to \gamma
A^0)$~\cite{Nason}, and $g_q$ is the effective Yukawa coupling to the
Higgs field to the up- or down-type quark-pair. In the NMSSM,
$g_c=\cos\theta_A/\tan\beta$ ($q=c$) for the charm quark and
$g_b=\cos\theta_A\tan\beta$ ($q=b$) for the bottom quark, where $\tan\beta$ is
the ratio of up- and down-type Higgs doublets, and $\cos\theta_A$
is the fraction of the nonsinglet component of the
$A^0$~\cite{Dermisek1, fayet}. The value of $\cos\theta_A$ is zero for a
pure $A^0$ singlet state~\cite{fayet1}. The branching
fraction of $J/\psi \to \gamma A^0$ is predicted to be in the range of
$10^{-9}-10^{-7}$ depending upon the $A^0$ mass and the NMSSM
parameters~\cite{Dermisek}. The branching fraction of $A^0 \to \mu^+\mu^-$ is predicted to be much larger for $\tan\beta \ge 1$~\cite{Dermisek1}. An experimental study of such a low-mass Higgs boson is desirable to test the SM~\cite{Lisanti} and to look for new physics beyond the SM~\cite{Hiller, Dermisek, fayet2}.

The BaBar~\cite{slacR1008}, CLEO~\cite{cleo}, and CMS~\cite{cms} experiment have searched for di-muon decays of $A^0$, and placed a strong exclusion upper limit on $g_b$. On the other hand, the BESIII measurements, sensitive on $g_c$, is complementary to those by considering $g_b$. The recent BESIII measurement~\cite{bes3-2}, based on 225
million $J/\psi$ events, is slightly lower than the BaBar
measurement~\cite{slacR1008} in the low-mass region for $\tan \beta
\le 0.6$. The combined measurements of the BESIII and BaBar 
have
revealed that the $A^0$ is mostly singlet in
nature because of obtained upper limit on $\cos\theta_A(=|\sqrt{g_bg_c}|)\times \sqrt{\mathcal{B}(A^0\to\mu^+\mu^-)}$, independent of $\tan\beta$, is very close to zero especially in the low-mass region~\cite{bes3-2}. However, this BESIII limit~\cite{bes3-2} is
still an order of magnitude above the theoretical
predictions~\cite{Dermisek}.  BESIII has recently accumulated about 39
times more $J/\psi$ events in comparison to the previous
measurement~\cite{bes3-2}, and these can be utilized to discover the
$A^0$ or exclude parameter space of the NMSSM~\cite{bes3jps}.

This paper describes the search for di-muon decays of a $CP$-odd light Higgs boson in radiative decays of $J/\psi$ using 9 billion $J/\psi$ events collected by the BESIII detector in 2009, 2018, and 2019~\cite{bes3jps}. Because muon particle identification (PID) was not available for the $J/\psi$ data collected in 2012, we exclude this data sample for the $A^0$ search.  

\section{BESIII Detector and Monte Carlo Simulation}

The BESIII detector~\cite{bes3det} records symmetric $e^+e^-$
collisions provided by the BEPCII storage
ring~\cite{Yu:IPAC2016-TUYA01}, which operates with a peak luminosity
of $1\times10^{33}$~cm$^{-2}$s$^{-1}$ in the center-of-mass (CM)
energy range from 2.0 to 4.95 GeV.  BESIII has collected large data
samples in this energy region~\cite{Ablikim:2019hff}. The cylindrical
core of the BESIII detector covers 93\% of the full solid angle and
consists of a helium-based multilayer drift chamber~(MDC), a plastic
scintillator time-of-flight system~(TOF), and a CsI(Tl)
electromagnetic calorimeter~(EMC), which are all enclosed in a
superconducting solenoidal magnet providing a 1.0~T magnetic
field. The solenoid is supported by an octagonal flux-return yoke with
resistive plate counter muon identification modules interleaved with
steel.  The MDC measures the momentum of charged particles with a
resolution of $0.5\%$ at 1 GeV/$c$. The EMC measures the photon
energies with a resolution of $2.5\%$ ($5\%$) at 1 GeV in the barrel
(end-cap) region. The time resolution of the TOF in the barrel region
is 68 ps. The time resolution of the TOF in the endcap region was 110
ps before 2015 and was improved to be 60 ps after upgrading with the
multi-gap resistive plate chambers. Muons with momentum above 0.5 GeV/c are identified by the iron flux return of the magnet instrumented with about 1272 $m^2$ of resistive plate muon counters (MUC) arranged in nine (eight) layers in the barrel (endcaps).

Simulated Monte Carlo (MC) events based on {\sc Geant4}~\cite{geant4}
are used to optimize the event selection criteria, to study the
potential backgrounds, and to determine the detector acceptance. A MC
sample of 9.0 billion inclusive $J/\psi$ events is used for the
background studies with the generic TopoAna tool~\cite{topo}. The known
$J/\psi$ decay modes are generated by the {\sc EvtGen}
generator~\cite{evtgen} with branching fractions taken from the
Particle Data Group (PDG)~\cite{pdg}, and the remaining unknown decay
modes by {\sc LUNDCHARM}~\cite{lundcharm}. The final state radiation
corrections are included in the MC simulation using {\sc
PHOTOS}~\cite{photos}. The production of the $J/\psi$ resonance
through $e^+e^-$ annihilation including the beam-energy spread and the
initial-state-radiation (ISR) is simulated by the {\sc
KKMC}~\cite{kkmc}. A $2.93$ fb$^{-1}$ $\psi(3770)$ data
sample~\cite{psipp,psipp1} is used to study the background from the
quantum electrodynamics (QED) process of $e^+e^- \to \gamma
\mu^+\mu^-$. To compute the detection efficiency, we generate 0.12
million simulated signal MC events at 23 different Higgs mass points
ranging from 0.212 to 3.0 GeV/$c^2$ with a phase-space model for the
$A^0 \to \mu^+\mu^-$ decay and a $P$-wave model for the $J/\psi \to
\gamma A^0$ decay~\cite{evtgen}.

\section{Data Analysis}
\label{section_selections}
 We select events with two oppositely charged tracks and at least one
 photon candidate. A photon candidate, reconstructed with clusters of
 energy deposited in the EMC, is selected with a minimum energy of 25
 MeV in the barrel region ($|\cos\theta| < 0.8$) or 50 MeV in the
 end-cap region ($0.86 < |\cos\theta| < 0.92$). The energy deposited
 in the nearby TOF is included to improve the energy resolution and
 reconstruction efficiency. The angle between a photon and the nearest
 extrapolated track in the EMC is required to be larger than 10
 degrees to remove bremsstrahlung photons. The EMC timing is required
 to be within 700 ns relative to the event start time to suppress
 electronic noise and energy deposits unrelated to the signal events.

Charged tracks are reconstructed from the ionization signals measured
by the MDC and are required to be in the MDC detection acceptance
region of $|\cos\theta|<0.93$, where $\theta$ is the angle of the
charged track with the $z$ axis, which is the axis of the MDC.
Further, their points of closest approach to the $z$-axis must be
within $\pm 10$ cm from the interaction point along the $z$ direction
and within $\pm 1$ cm in the plane perpendicular to $z$. To suppress
contamination by electrons and pions, both charged tracks are
required to satisfy the following selection criteria: 1) $E_{\rm
cal}^{\mu}/p < 0.9$ $c$, 2) $0.1 < E_{\rm cal}^{\mu} < 0.3$ GeV, and
3) the absolute value of the time difference between the TOF and
expected muon time ($\Delta t^{\rm TOF}$) must be less than 0.26
ns. Here, $E_{\rm cal}^{\mu}$ is the energy deposited in the EMC by the
$\mu^+/\mu^-$ particle, and $p$ is the momentum of the
charged muon track. To further improve the purity of muons, one of the
charged tracks is required to have its penetration depth in the MUC be
greater than ($-40.0+70\times p/({\rm GeV}/c)$) cm for $0.5 \le p \le
1.1$ GeV/$c$ and 40 cm for $p>1.1$ GeV/$c$.

The two muon tracks are required to originate from a common vertex by
performing a vertex fit to form an $A^0$ candidate. A four-constraint
(4C) kinematic fit is performed with the two charged tracks and one of
the photon candidates in order to improve the mass resolution of the
$A^0$ candidate. If there is more than one $\gamma \mu^+\mu^-$
candidate, the candidate with the minimum value of the $\chi^2$ from
the 4C kinematic fit ($\chi_{4 \rm C}^2$) is selected, and the $\chi_{4
\rm C}^2$ is required to be less than 40 to reject backgrounds
from $J/\psi \to \pi^+\pi^-\pi^0$ and $e^+e^- \to \gamma
\pi^+\pi^-\pi^0$. We reject fake photons by requiring the di-muon
invariant mass ($m_{\mu^+\mu^-}$) obtained from the 4C kinematic fit
to be less than 3.04 GeV/$c^2$. To suppress backgrounds from
$e^+e^-\to \gamma \mu^+\mu^-$ and $J/\psi \to \mu^+\mu^-(\gamma)$, the
absolute value of the cosine of the muon helicity angle
($\cos\theta_{\mu}^{\rm hel}$), defined as the angle between the
direction of one of the muons and the direction of the $J/\psi$ in the
$A^0$ rest frame, is required to be less than 0.92.

\begin{figure} \centering
\includegraphics[width=0.5\textwidth]{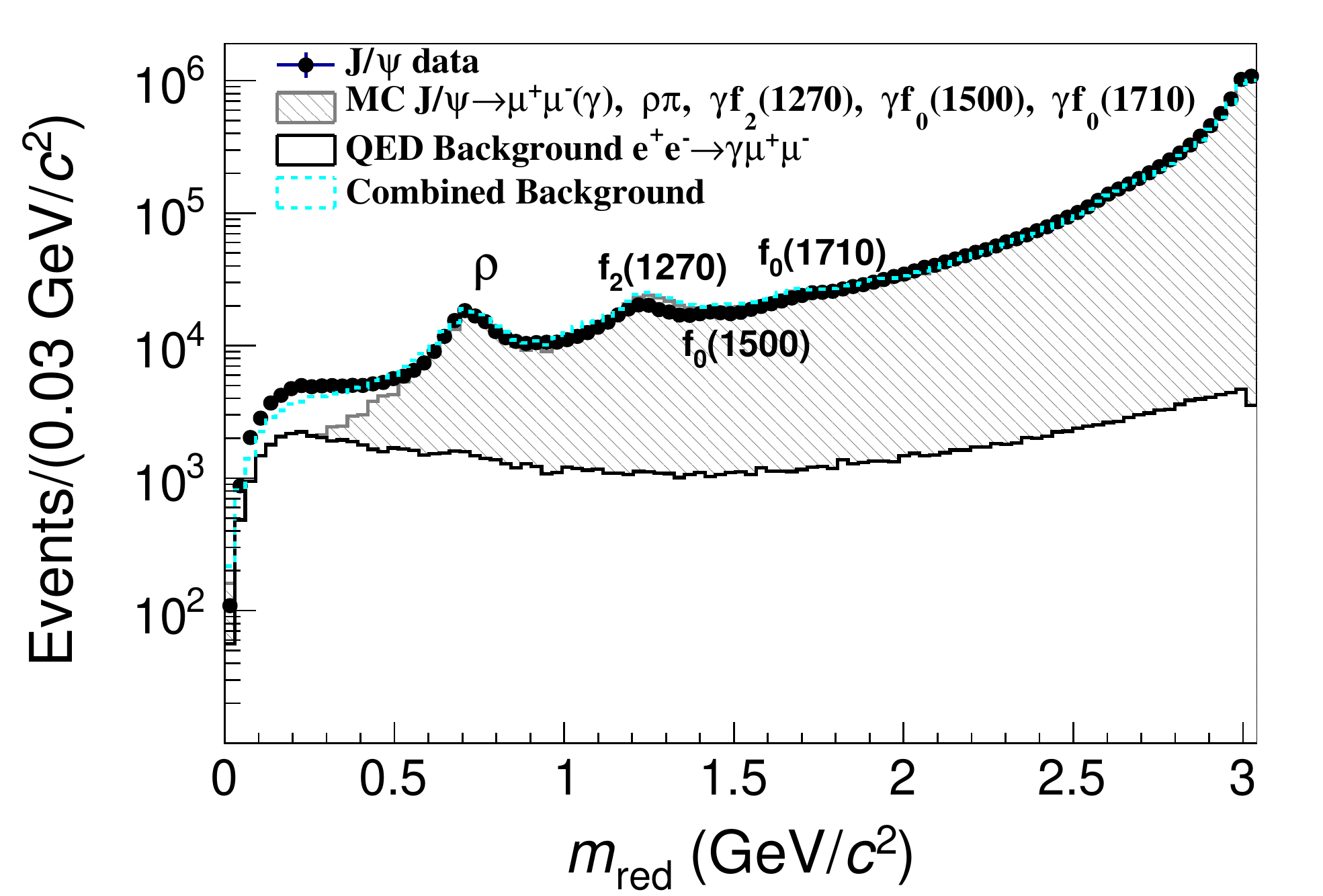} \caption{The $m_{\rm
red}$ distribution of data (black dot points with error bars),
together with the background predictions of the QED $e^+e^- \to \gamma
\mu^+ \mu^-$ process from $\psi(3770)$ data (black histogram) and
$J/\psi \to \rho \pi$, $\mu^+\mu^-(\gamma)$, $\gamma f$ ($f=f_2(1270),
f_0(1500), f_0(1710)$) decays from MC samples of
those processes (gray pattern histogram). The dashed cyan
histogram represents the combined background. } \label{mred}
\end{figure}

After the above selection criteria, we determine the signal yield as a function of $m_{A^0}$ in the interval
of $0.212 \le m_{A^0} \le 3.0$ GeV/$c^2$ by performing a series of
one-dimensional unbinned extended maximum likelihood (ML) fits to the reduced mass, $m_{\rm red}=\sqrt{m_{\mu^+\mu^-}^2-4m_{\mu}^2}$ distribution of surviving events.  Fig.~\ref{mred} shows the $m_{\rm red}$ distribution of surviving events together
with the background predictions from various simulated MC samples and
$2.93$ fb$^{-1}$ of $\psi(3770)$ data~\cite{psipp,psipp1}. We use
$m_{\rm red}$ rather than $m_{\mu^+\mu^-}$ because it is easier to
model the non-peaking background across the entire $m_{A^0}$ region,
in particular, the kinematic threshold region
$m_{\mu^+\mu^-}\approx 2m_{\mu}$ ($m_{\rm red} \approx 0$).  The
non-peaking background is dominated by $e^+e^- \to \gamma \mu^+\mu^-$
and $J/\psi \to \mu^+\mu^-(\gamma)$, and the peaking background by
$J/\psi \to \rho/\omega \pi$ and $J/\psi \to \gamma f$
$(f=f_2(1270),f_0(1500),f_0(1710))$ decays, where both $\rho/\omega$
and $f$ decay to $\pi^+\pi^-$. The $m_{\rm red}$ distribution of data is generally well described by the background predictions,
except in the low-mass region, where {\sc KKMC}~\cite{kkmc} fails to reproduce the ISR events for the $e^+e^- \to \gamma J/\psi, J/\psi \to \mu^+ \mu^-$ process.
This disagreement has little impact on the search because the signal extraction procedure does not depend on the background predictions. 

 The fit function includes the
contributions of signal, continuum background and peaking background
components from $\rho/\omega$, $f_2(1270)$, $f_0(1500)$, and
$f_0(1710)$ mesons. Table~\ref{fitrange} summarizes the $m_{\rm red}$
fit intervals for the various $m_{A^0}$ points used to handle both
non-peaking and peaking backgrounds smoothly.

\begin{table}
\centering
\caption{The $m_{\rm red}$ fit intervals for various $m_{A^0}$ points. }
\begin{tabular}{p{2.5cm} |p{1.75cm}|p{2.5cm}}
        \hline    \hline

  $m_{\rm red}$ fit interval (GeV/$c^2$) & $m_{A^0}$ points  (GeV/$c^2$) & Order of Polynomial function \\
        \hline \hline
 $0.002 - 0.45$    & $0.212,  0.4$ & $5^{th}$ \\
 $0.3 -  0.65$    & $0.401, 0.6$   & $4^{th}$ \\
 $0.4 -  1.06$    & $0.601, 1.0$   & $3^{rd}$ \\
 $0.95 - 1.95$    & $1.001, 1.8$   & $2^{nd}$ \\
 $1.7 -  2.5$     & $1.802, 2.4$   & $5^{th}$ \\
 $2.3 -  2.7$     & $2.402, 2.6$   & $4^{th}$ \\
 $2.5 -  2.9$     & $2.602, 2.848$ & $5^{th}$ \\
 $2.75 - 3.0$     & $2.85,  2.90$  & $6^{th}$ \\         
 $2.85 - 3.032$   & $2.902,  3.0$  & $5^{th}$ \\   \hline \hline  
\end{tabular}
\label{fitrange}
\end{table}

Simulated MC samples are used to develop the probability density
functions (PDFs) of signal and backgrounds. 
The $A^0$ is assumed to be a scalar or pseudo-scalar particle with a very narrow decay width in comparison to the experimental resolution~\cite{Res}.
We describe the $m_{\rm red}$ distribution of the signal PDF by the sum of two Crystal Ball
(CB) functions~\cite{CB} with a common peak value and opposite side tails. The $m_{\rm red}$
resolution varies from $2$ MeV/$c^2$ to $12$ MeV/$c^2$ while the
signal efficiency varies between 27\% and 53\% depending upon the muon
momentum values at different $A^0$ mass points. We interpolate the
signal efficiency and signal PDF parameters linearly between the mass
points of the generated signal MC events. The non-peaking background
PDF is described by a function $\tanh(\sum_{l=1}^5p_lm_{\rm red}^l)$
in the threshold mass region of $0.212 \le m_{A^0} \le 0.40$
GeV/$c^2$, where $p_l$ are the polynomial coefficients. This function
provides a threshold like behavior in the low-mass region of the $m_{\rm
red}$ distribution and passes through the origin when $m_{\rm
red}=0$. In the other mass regions, we use second, third, fourth,
fifth or sixth-order Chebyshev polynomial function to describe the non-peaking
background PDFs detailed in Table~\ref{fitrange}.  We determine the initial parameters of these
background PDFs using a cocktail MC sample of all possible non-peaking
backgrounds to achieve better agreement between data and the fit models.

To take into account the well-known structure of the $\rho$-$\omega$
interference, we describe the peaking background PDF of the $m_{\rm
red}$ distribution with the Gounaris and Sakurai (GS) function in the
range of $0.4 \le m_{\rm red} \le 1.06$ GeV/$c^2$~\cite{gsmodel}. The
fit formula, detailed in Ref.~\cite{BaBar_model}, is the same as that
used previously by the BaBar~\cite{BaBar_model} and
BESIII~\cite{psipp1} experiments in the measurement of the $e^+e^- \to
\pi^+\pi^-$ cross-section in the $\rho/\omega$ mass region.  The
amplitudes for the higher $\rho$ states, $\rho(1450)$, $\rho(1700)$,
and $\rho(2150)$, as well as the massses and widths of those states
are taken from Ref.~\cite{BaBar_model}. We fix the $\omega$ width
according to the PDG~\cite{pdg} value and float the other
parameters during the fit. We describe the peaking background PDFs
corresponding to $f_2(1270)$, $f_0(1500)$ and $f_0(1710)$ resonances
by the sum of the two CB functions~\cite{CB} using the parameters
determined from MC samples of $J/\psi \to \gamma f$, $f \to
\pi^+\pi^-$ decays, where $f = f_2(1270)$, $f_0(1500)$, and
$f_0(1710)$ mesons.

\begin{figure}
  \centering \includegraphics[width=0.5\textwidth]{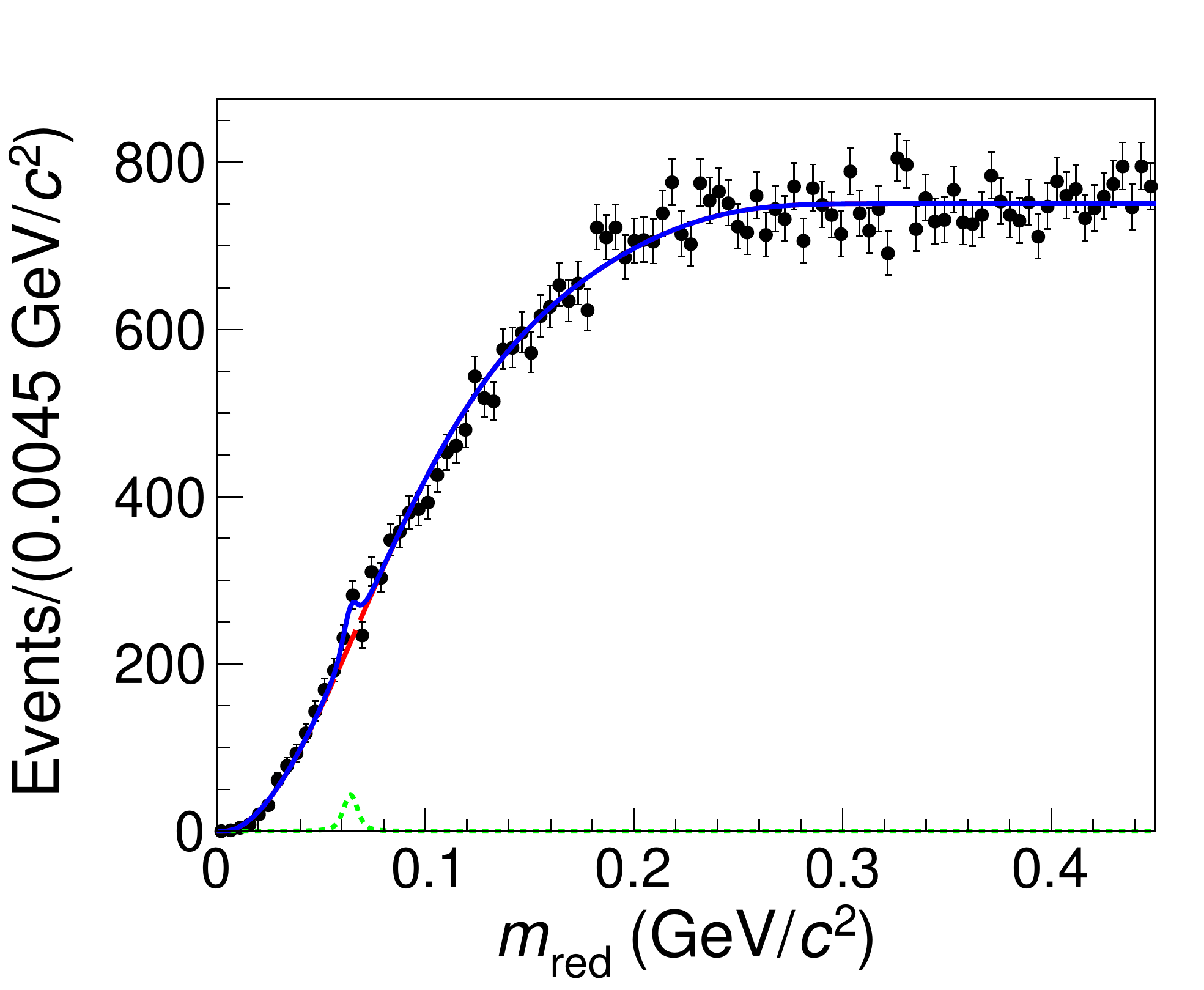}
  \includegraphics[width=0.5\textwidth]{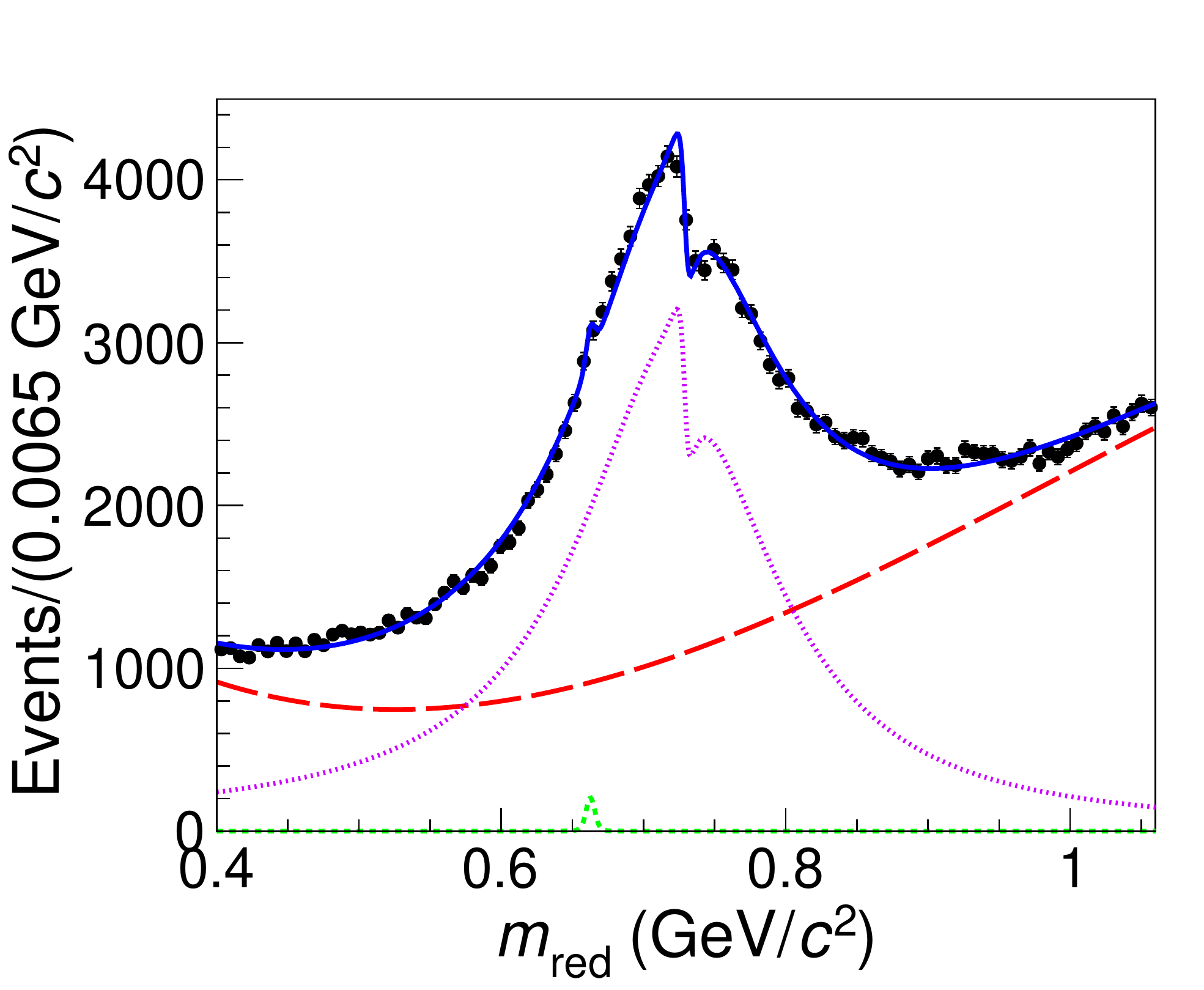} \caption{Fits to
  the $m_{\rm red}$ distributions for (top) $m_{A^0}=0.221$ GeV/$c^2$
  and (bottom) $m_{A^0} = 0.696$ GeV/$c^2$. The corresponding local
  significance values at these mass points are observed to be
  $3.3\sigma$ and $3.5\sigma$, respectively. Black dots with error
  bars represent the data, the red long-dashed curve the non-peaking
  background, the pink dotted curve the peaking background, the green dashed
  curve the signal PDF, and the solid blue curve the total fit
  results.  In the bottom figure, the well-known $\rho-\omega$
  interference is taken care of by describing the peaking background
  PDF of the $m_{\rm red}$ distribution by a GS
  function~\cite{gsmodel, BaBar_model}, as described in the
  text.}  \label{Proj} \end{figure}

The search for the $A^0$ narrow resonance is performed in steps of
approximately half the $m_{\rm red}$ resolution, i.e., 1 MeV/$c^2$
in the mass range of $0.22 \le m_{A^0} \le 1.5$ GeV/$c^2$ and 2.0
MeV/$c^2$ in the other $m_{A^0}$ regions, with a total of 2,035
$m_{A^0}$ points. The PDF parameters of the signal and peaking
backgrounds of $J/\psi \to \gamma f$ are fixed while the non-peaking
background PDF, and the numbers of the signal, peaking, and non-peaking
background events are floated. Plots of the fit to
the $m_{\rm red}$ distribution for two selected mass points are shown
in Fig.~\ref{Proj}.

The fit is repeated with alternative signal, peaking, and
non-peaking background PDFs to determine the systematic uncertainties for
the numbers of signal events associated with the corresponding PDFs at
each $m_{A^0}$ point. The uncertainty associated with the signal PDF
is studied by replacing the sum of the two CB functions with a \rm{\lq
Cruijff\rq} function~\cite{Cruijff}. The uncertainty associated with
the $\rho-\omega$ peak is evaluated by varying the $\rho$ and $\omega$
contributions in the formula of Eq.(26) of
Ref.~\cite{BaBar_model}. The uncertainty due to the peaking background
of $J/\psi \to \gamma f$ is studied by replacing the sum of the two CB
functions with the simulated MC samples of the corresponding decay
processes convolved with a Gaussian function whose parameters are
floated during the fit. The uncertainty due to the non-peaking
background PDF is studied by replacing the
$\tanh(\sum_{l=1}^5p_lm_{\rm red}^l)$ and $n^{th}$ order Chebyshev
polynomial function with $\tanh(\sum_{l=1}^6p_lm_{\rm red}^l)$ and
$(n+1)^{th}$ order Chebyshev polynomial functions, respectively, in
the fit. The one with the largest signal yield among these fit scenarios is
considered to produce the final result.

\begin{figure} \centering
\includegraphics[width=0.5\textwidth]{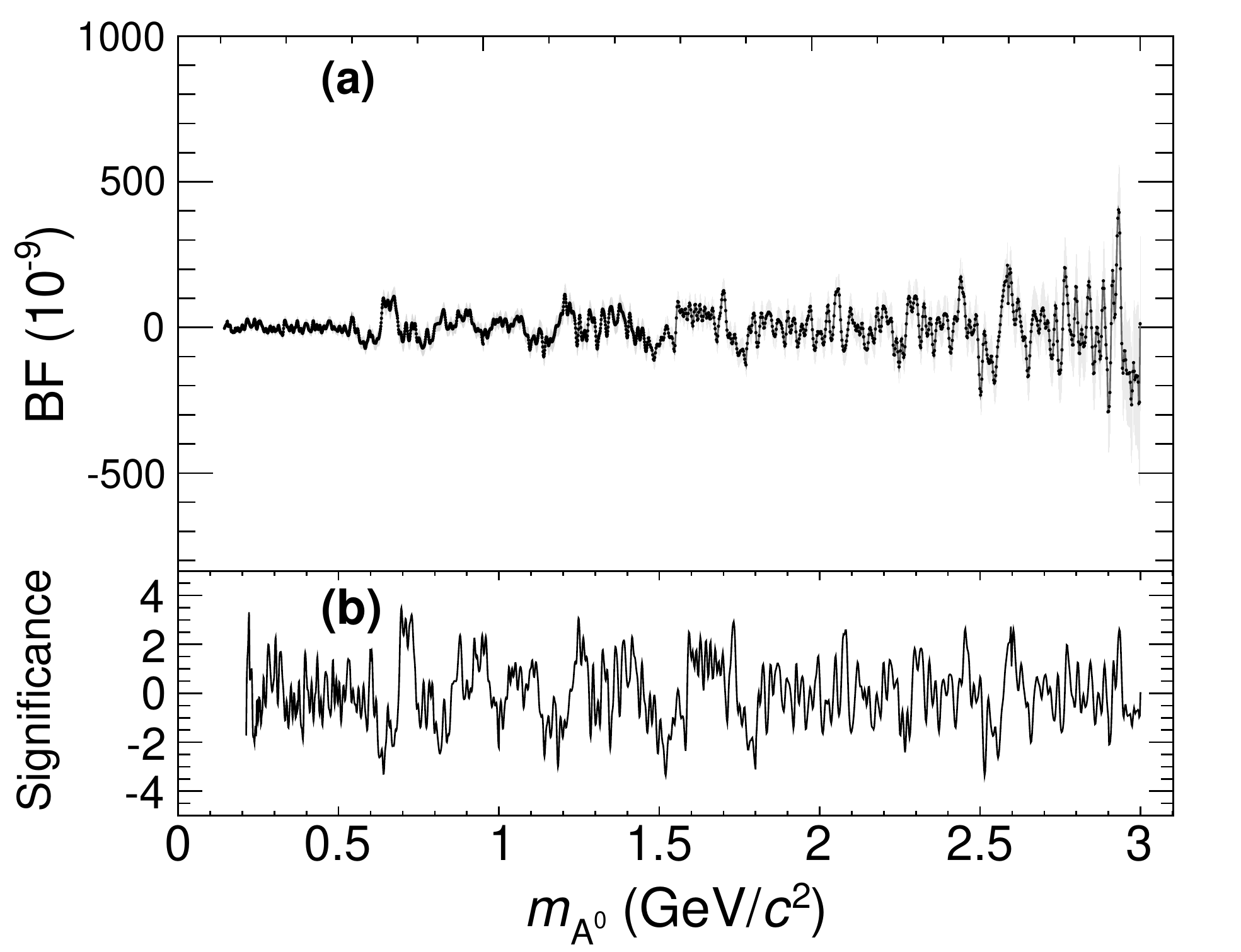} \caption{(a) The
product branching fractions $\mathcal{B}(J/\psi \to \gamma A^0) \times
\mathcal{B}(A^0 \to \mu^+\mu^-)$ (BF) and (b) signal significance
($\mathcal{S}$) obtained from the fit as a function of $m_{A^0}$.}
\label{yield} \end{figure}

The product branching fraction of $J/\psi \to \gamma A^0$ and $A^0 \to \mu^+\mu^-$ as a function of $m_{A^0}$ is calculated as

\begin{equation}
\mathcal{B}(J/\psi \to \gamma A^0) \times \mathcal{B}(A^0 \to
\mu^+\mu^-) =\frac{N_{sig}}{\epsilon \cdot N_{J/\psi}},
\label{Eq:BF}
\end{equation}

\noindent where $N_{sig}$ is the number of signal events, $\epsilon$
is the signal selection efficiency, and $N_{J/\psi}=(8.998 \pm
0.039)\times 10^9$ is the number of $J/\psi$ events.
Fig.~\ref{yield} shows the plots of the product
branching fractions $\mathcal{B}(J/\psi \to \gamma A^0) \times
\mathcal{B}(A^0 \to \mu^+\mu^-)$ and the statistical significance,
defined as $\mathcal{S}={\rm sign}(N_{sig})\sqrt{2{\rm
ln}(\mathcal{L}_{\rm max}/\mathcal{L}_0)}$, where $\mathcal{L}_{\rm
max}~(\mathcal{L}_0)$ is the maximum likelihood value for a fit with
number of signal events being floated (fixed at zero). The largest
upward local significance value is determined to be $3.5\sigma$ at
$m_{A^0}=0.696$ GeV/$c^2$. Based on a large ensemble of pseudo
experiments~\cite{slacR1008}, the probability of observing a fluctuation of
$\mathcal{S}\ge 3.5\sigma$ is estimated to be $12\%$. The
corresponding global significance value is determined to be at the
level of $1\sigma$. Thus, we conclude that no evidence of Higgs
production is found within the searched $m_{A^0}$ regions. 

\section{Systematic Uncertainties} \label{syst} According to
Eq.~\ref{Eq:BF}, the systematic uncertainties for the branching
fraction measurement include those from the number of signal events,
the reconstruction efficiency, and the number of $J/\psi$ events. 
The uncertainties associated with the number of signal events originating
from the PDF parameters of signal and backgrounds are considered by
performing alternative fits at each $m_{A^0}$ point. 

Pseudo
experiments are utilized to test the reliability of the fit
procedures and compute the fit bias, which may appear due to imperfect
signal and background modeling. The same fit procedure is performed in
each pseudo experiment. The resultant average difference between the
input and output signal yields is determined to be 0.3 events. We
consider it as an additive systematic uncertainty ($\sigma_{\rm
add}$), which may affect the significance of any observation but does
not scale with the reconstructed signal yield.

The uncertainties
associated with the reconstruction efficiency and the number of
$J/\psi$ events don't affect the significance of any
observation. Thus, we consider them as multiplicative systematic
uncertainties ($\sigma_{\rm mult}$) and scale with the number of
reconstructed signal events. The uncertainty associated with the
reconstruction efficiency includes those from tracking, PID, and
the photon detection efficiency.

The uncertainty due to MDC tracking is determined to be $1\%$ per
track using the high statistics control samples of $J/\psi \to \rho
\pi$ and $J/\psi \to p\bar{p}\pi^+\pi^-$. A total of $2.0\%$
systematic uncertainty is assigned for the two charged tracks in this
analysis. The systematic uncertainty associated with the photon
reconstruction efficiency is determined using a control sample of
$e^+e^- \to \gamma \mu^+\mu^-$ in which the ISR photon is predicted
using the four momenta of the two charged tracks. This sample also
includes the dominant contribution from $J/\psi \to \gamma \pi^+\pi^-$
decay, including all the possible intermediate resonances. The
relative difference in efficiency between data and MC is found to be
$0.2\%$, which is considered as a systematic uncertainty.

A control sample of $J/\psi \to \mu^+\mu^-(\gamma)$ is used to
evaluate the systematic uncertainty due to the muon PID,
$\cos\theta_{\mu}^{\rm hel}$, and $\chi_{4 \rm C}^2$ requirements. In
this sample, one track is tagged with a tight muon PID. The final
uncertainty associated with the muon PID also takes into account the
fraction of events with one or two tracks identified as muons obtained
from the simulated signal MC sample. The corresponding uncertainties,
computed as the relative change in efficiency between data and MC, are
determined to be $(2.9-4.1)\%$, $0.8\%$ and $1.8\%$, respectively. The
systematic uncertainty due to the number of $J/\psi$ events is
$0.3\%$ using $J/\psi$ inclusive hadronic events.
Table~\ref{Systematic} summarizes the fit bias and multiplicative
sources of the systematic uncertainties, where we obtain the total
$\sigma_{\rm mult}$ by adding the individual ones in quadrature. The
total $\sigma_{\rm mult}$ varies between $4.1\%$ to $5.0\%$ depending on the Higgs mass point.  The final systematic uncertainty is
calculated as $\sqrt{\sigma_{\rm add}^2+(\sigma_{\rm mult}\times
  N_{sig})^2}$.

\begin{table}
\centering
\caption{The fit bias and multiplicative sources of the systematic
  uncertainties. The systematic uncertainties associated with the
  signal, peaking, and non-peaking background PDFs are taken as the
  largest difference of signal yield among the alternative fit
  scenarios at each $m_{A^0}$ point as described in
  Sect.~\ref{section_selections}.}  \centering
\begin{tabular}{c |c}
        \hline \hline
Source  &  Uncertainty  \\
\hline 
\multicolumn{2}{c}{Additive systematic uncertainties (events) } \\
\hline
 Fit Bias                     & 0.3                \\

Total                         & 0.3    \\

\hline
\multicolumn{2}{c}{Multiplicative systematic uncertainties ($\%$)} \\
\hline

Tracking                            & 2.0       \\ 
Photon detection efficiency         & 0.2     \\
Depth in MUC                        & 2.9 -- 4.1     \\
$E_{cal}^{\mu}$                       & 0.1     \\ 
$\Delta t^{ \rm TOF}$                     & Negl.     \\
Cos$\theta_{\mu}^{hel}$               & 0.8     \\ 

$\chi_{4 \rm C}^2$                        & 1.8   \\ 
$J/\psi$ counting                   & 0.7  \\ \hline 
Total                            & 4.1 -- 5.0     \\ 
\hline
\end{tabular}
\label{Systematic}
\end{table}

\section{Result}
\begin{figure}
\centering
\includegraphics[width=0.5\textwidth]{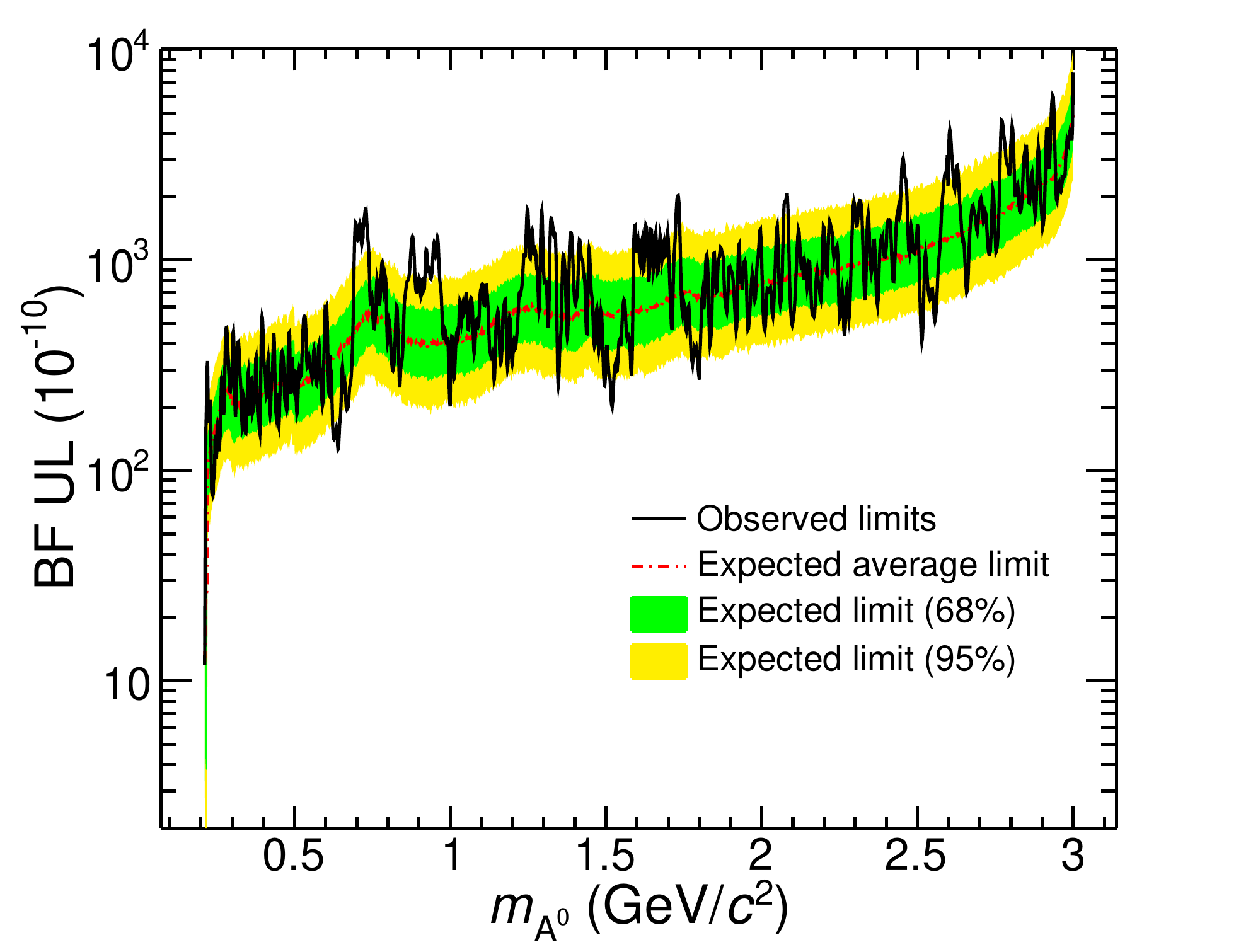}
\caption{The $90\%$ C.L. upper limits on the product branching
  fractions $\mathcal{B}(J/\psi \to \gamma A^0)\times \mathcal{B}(A^0
  \to \mu^+\mu^-)$ versus $m_{A^0}$ including all the uncertainties,
  together with the expected limits computed using a large number of
  pseudo experiments. The inner and outer bands correspond to $68\%$
  and $95\%$ of the expected limit values and include the statistical
  uncertainties only. }
\label{ulBF}
\end{figure}

Since no evidence of Higgs production is found, we set $90\%$
confidence level (C.L.) upper limits on the product branching fractions
$\mathcal{B}(J/\psi \to \gamma A^0)\times \mathcal{B}(A^0 \to
\mu^+\mu^-)$ as a function of $m_{A^0}$ using a Bayesian
method~\cite{pdg} after incorporating the systematic uncertainty by
smearing the likelihood curve with a Gaussian function having a width
equal to the systematic uncertainty. The limits vary in the range of
$(1.2-778.0)\times 10^{-9}$ for the Higgs mass region of $0.212 \le
m_{A^0} \le 3.0$ GeV/$c^2$ depending on the $m_{A^0}$ point, as shown
in Fig.~\ref{ulBF}. The new measurement has a 6-7 times improvement
over the previous BESIII measurement~\cite{bes3-2}.

\begin{figure}
\centering
\includegraphics[width=0.5\textwidth]{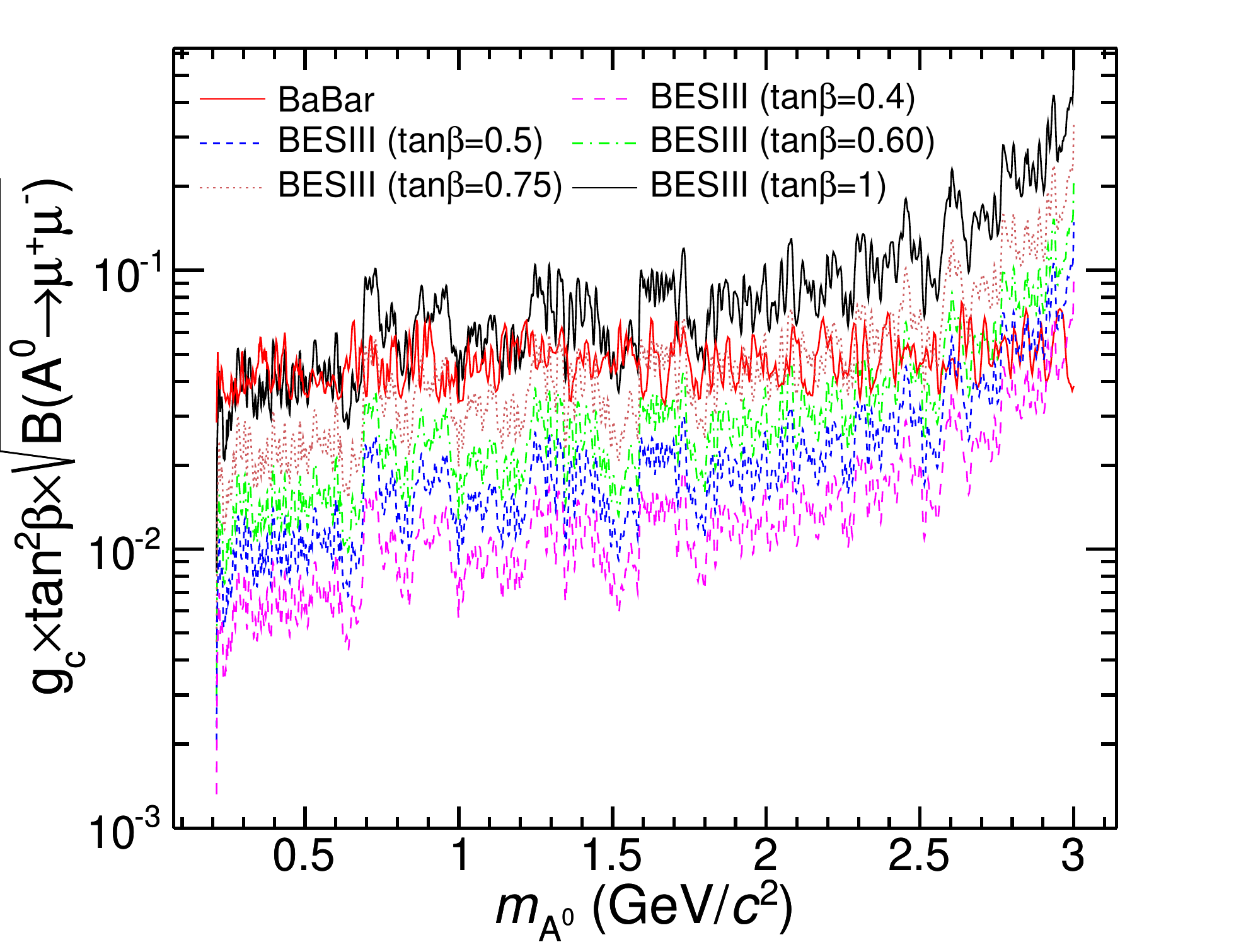}
\caption{The $90\%$ C.L. upper limits on the effective Yukawa coupling of
  the Higgs field to the bottom-quark pair $g_b(=g_c\tan^2\beta)\times
  \sqrt{\mathcal{B}(A^0 \to \mu^+\mu^-)}$ as a function of $m_{A^0}$
  for different values of $\tan\beta$ together with the BaBar
  measurement. Our results are slightly better than the BaBar measurement
  in the low mass region for $\tan\beta =1$.}
\label{nmssm}
\end{figure}

To compare our results with the BaBar measurement~\cite{slacR1008}, we
also compute $90\%$ C.L. upper limits on the effective Yukawa coupling of
the Higgs fields to the bottom-quark pair $g_b(=g_c\tan^2\beta)\times
\sqrt{\mathcal{B}(A^0 \to \mu^+\mu^-)}$ as a function of $m_{A^0}$ for
different values of $\tan\beta$ using Eq.(\ref{yukawa}) as shown in 
Fig.~\ref{nmssm}. Our new measurement is slightly better than the
BaBar measurement~\cite{slacR1008} in the low-mass region for
$\tan\beta = 1.0$.

\section{Summary}
We search for di-muon decays of $A^0$ in $J/\psi \to \gamma A^0$ using
$9.0$ billion $J/\psi$ events collected by the BESIII detector. No
evidence of Higgs production is found, and we set $90\%$ C.L. upper
limits on product branching fractions $\mathcal{B}(J/\psi \to \gamma
A^0)\times \mathcal{B}(A^0 \to \mu^+\mu^-)$ in the range of
$(1.2-778.0)\times10^{-9}$ for $0.212 \le m_{A^0} \le 3.0$
GeV/$c^2$. This result has an improvement by a factor of 6-7 over the
previous BESIII measurement~\cite{bes3-2}, and is better than the
BaBar measurement~\cite{slacR1008} for $m_{A^0} \le 0.7$ GeV/$c^2$ for
$\tan\beta =1$. Thus, our measurement is more stringent for $m_{A^0} \le 0.7$ GeV/$c^2$ over the existing experimental results~\cite{slacR1008, cleo, cms, bes3-2,bes3-1}. The new BESIII limit is also lower than the
theoretical prediction at the threshold Higgs mass point of $0.212$
GeV/$c^2$, and thus constrains a large fraction of the parameter space
of the new physics models, including NMSSM~\cite{Hiller,
  Dermisek, fayet2}.

\section{Acknowledgments}
The BESIII collaboration thanks the staff of BEPCII, the IHEP computing center and the supercomputing center of USTC for their strong support. This work is supported in part by National Key R$\&$D Program of China under Contracts Nos. 2020YFA0406300, 2020YFA0406400; National Natural Science Foundation of China (NSFC) under Contracts Nos. 11335008, 11625523, 11625523, 11635010, 11735014, 11822506, 11835012, 11935015, 11935016, 11935018, 11961141012, 12022510, 12025502, 12035009, 12035013, 12035013, 12061131003, 11705192, 11950410506, 12061131003; the Chinese Academy of Sciences (CAS) Large-Scale Scientific Facility Program; Joint Large-Scale Scientific Facility Funds of the NSFC and CAS under Contracts Nos. U1732263, U1832207, U1832103, U2032111; CAS Key Research Program of Frontier Sciences under Contract No. QYZDJ-SSW-SLH040; 100 Talents Program of CAS; INPAC and Shanghai Key Laboratory for Particle Physics and Cosmology; ERC under Contract No. 758462; European Union Horizon 2020 research and innovation programme under Contract No. Marie Sklodowska-Curie grant agreement No 894790; German Research Foundation DFG under Contracts Nos. 443159800, Collaborative Research Center CRC 1044, FOR 2359, FOR 2359, GRK 214; Istituto Nazionale di Fisica Nucleare, Italy; Ministry of Development of Turkey under Contract No. DPT2006K-120470; National Science and Technology fund; Olle Engkvist Foundation under Contract No. 200-0605; STFC (United Kingdom); The Knut and Alice Wallenberg Foundation (Sweden) under Contract No. 2016.0157; The Royal Society, UK under Contracts Nos. DH140054, DH160214; The Swedish Research Council; U. S. Department of Energy under Contracts Nos. DE-FG02-05ER41374, DE-SC-0012069.




\begin{thebibliography}{99}
\bibitem{Cruijff}  The Cruijff function is an asymmetric Gaussian function having a common mean ($\mu$),  different left-right resolutions ($\sigma_{L,R}$) and non-Gaussian tails ($\alpha_{L,R}$): $f(x)={\rm exp}((x-\mu)^2/(2\sigma_{L,R}^2+\alpha_{L,R}(x-\mu)^2))$.
\end{thebibliography}
\end{document}